# Assessing the Limits of Graph Neural Networks for Vapor-Liquid Equilibrium Prediction: A Cryogenic Mixture Case Study

Aryan Gupta
arygupta@umich.edu

## Abstract

Accurate and fast thermophysical models are needed to embed vapor-liquid equilibrium (VLE) calculations in design, optimization, and control loops for cryogenic mixtures. This study asks whether a structure-aware graph neural network (GNN; DimeNet++) trained on GERG-2008/CoolProp data can act as a practical surrogate for an equation of state (EoS). We generate a ternary dataset over 90-200 K and pressures to 100 bar, curate it with a 15% density filter (reducing 5,200 states to 1,516), and pair each state with a lightweight molecular-dynamics snapshot to supply structural features. The model is trained in two stages; pretraining on residual Helmholtz energy followed by pressure fine-tuning with a stability penalty; and evaluated via single-phase interpolation tests, solver-free derivative-quality diagnostics, an audited VLE driver, and a latency benchmark. Within its regime, the GNN interpolates single-phase properties reasonably well; however, the VLE driver accepts no GNN equilibria on tested binaries (all plotted VLE points are CoolProp fallback or the solver fails), and diagnostic probes reveal jagged $P(V|T)$ paths and thermal-stability flags concentrated in dense/cold regions, indicating insufficient derivative smoothness/consistency for robust equilibrium solving. An end-to-end timing comparison shows no single-phase speed advantage relative to CoolProp (tens of milliseconds vs sub-millisecond). We conclude that, as configured, the surrogate in this study is not solver-ready for VLE and offers no runtime benefit; its value is methodological, delineating failure modes and pointing to remedies such as physics-informed training signals and targeted coverage near phase boundaries.

## 1. Introduction

The cryogenic ternary mixture $CO_2 - CH_4 - N_2$ appears across natural-gas conditioning, carbon-management strategies, and low-temperature process operations where accurate phase behavior and property predictions determine safety margins, energy use, and separation performance [1]. These applications demand models that are both fast (so they can be embedded inside design loops, optimizers, and controllers) and accurate across the single-phase regions and, crucially, at vapor-liquid equilibrium (VLE) where small derivative errors can cause large composition or pressure mismatches [2]. This study's aim is to examine whether a modern, structure-aware machine-learning surrogate can satisfy that dual requirement for cryogenic $CO_2 - CH_4 - N_2$.

Classical thermodynamic modeling offers a spectrum of trade-offs. Cubic equations of state (EoS) remain popular because they are analytic, fast, and easy to differentiate; however, they can exhibit substantial bias for non-ideal mixtures and at cryogenic conditions, especially when extrapolating beyond parameter-fitting regimes [3]. At the other end, high-fidelity multi-parameter models and first-principles simulations provide improved accuracy and richer physics but at significant computational cost and implementation complexity [4,5]. For $CO_2 - CH_4 - N_2$, the GERG-2008 family of models accessed via CoolProp offers a strong reference baseline but is not designed as a learnable, structural surrogate; molecular dynamics (MD) offers EoS-independent validation but is far too slow for routine process use [5,6]. This landscape motivates a data-driven alternative that might capture non-ideal behavior with near-simulation accuracy while retaining millisecond-level evaluation speed.

This study investigates a graph neural network (GNN) surrogate built on DimeNet++, trained to map thermodynamic state and molecular structure to residual properties. The training dataset is generated from GERG-2008 via CoolProp across cryogenic temperatures and pressures (90-120 K; up to 200 bar) and the full ternary composition space, then curated using a physically motivated 15% density rule to remove spurious "liquid" points, yielding a final set on the order of 1,516 states from an initial 5,200 [6,7]. For each curated state, a short MD relaxation provides a structure snapshot (atomic numbers $z$ and coordinates $pos$) so the model can learn how composition and local geometry interact with temperature and molar volume to shape non-ideal behavior [8,9]. The model is trained in two stages: (i) a 50-epoch pretraining on residual Helmholtz energy, and (ii) a 100-epoch fine-tuning on pressure only (Huber loss) with a $C_V$ stability penalty; $U$ is reconstructed from $A_{res}$ rather than directly trained [10].

**Problem statement:** Can a GNN surrogate trained on curated GERG-2008/CoolProp data and MD-derived structures serve as a practical, thermodynamically reliable stand-in for an equation of state (EoS) for $CO_2 - CH_4 - N_2$ at cryogenic conditions, accurate for single-phase properties and sufficiently smooth and consistent to drive a derivative-based VLE solver? This study adopts a cautious, evidence-first approach: quantify single-phase interpolation accuracy, probe outlier regimes, and then test whether the learned surface possesses the derivative quality needed for equilibrium calculations.

**Thesis and objectives:** This study's thesis is that a structure-aware GNN can learn a useful approximation to the residual property landscape for cryogenic $CO_2 - CH_4 - N_2$ but, without explicit thermodynamic-consistency enforcement or targeted two-phase training signals, may not produce the smooth, self-consistent derivatives that VLE solvers require. Accordingly, this study's objectives are fourfold: (1) construct the curated dataset and structural inputs; (2) develop and train a DimeNet++-based surrogate to predict $A_{res}$ from state and structure and reconstruct $P, U$ from it; (3) evaluate single-phase accuracy with standard diagnostics; and (4) investigate whether the surrogate can support VLE, first directly, then in a decoupled hybrid assessment that isolates derivative behavior at known equilibrium states [11,12].

**Preview of contributions:** First, a cryogenic $CO_2 - CH_4 - N_2$ dataset is curated and a DimeNet++-based Helmholtz-energy surrogate that achieves accurate single-phase interpolation within its regime is trained. Second, a tiered VLE driver is audited and no accepted GNN equilibria is found on the tested binaries and temperatures; all plotted VLE points are CoolProp fallback and remaining grid points that lack a reference value are logged as "Solver Failed". Third, solver-free diagnostics that identify derivative/stability issues concentrated in dense/cold regions are provided. Finally, a single-phase latency benchmark showing no speed advantage of the current surrogate over a GERG-2008 baseline is reported [10,13,14,15,16].

**This study is structured as follows:** Section 2 describes data generation and curation; Section 3 the model and inputs; Section 4 the training protocol; Section 5 single-phase interpolation results; Section 6 the audited VLE attempts; Section 7 derivative-quality diagnostics; Section 8 latency; Section 9 discussion; and Section 10 conclusions.

## 2. Data: Dataset Generation & Curation

**Source and scope:** The primary training data were generated from the GERG-2008 family of models accessed through the CoolProp library, providing a consistent reference for thermodynamic properties over the cryogenic regime relevant to $CO_2 - CH_4 - N_2$ mixtures [6,17]. The sampling space spans temperatures from 90-120 K, pressures up to 200 bar, and the full ternary composition simplex so that both single-component limits and interior mixture states are represented [18]. From this design, an initial set of 5,200 state points was produced, of which 3,312 state points were eliminated due to unavailability of verifiable data in the CoolProp library, bringing the total number of points down to 1,516.

**Curation via density filter:** To remove physically implausible "liquid" states and improve data quality near regions where surrogate models are most fragile, this study applied a domain-informed density-based filter: points failing a 15% density criterion between CoolProp data and MD data were discarded. This procedure reduced the training set to 1,516 curated states while preserving broad coverage of temperature, pressure, and composition needed for interpolation within single-phase regions [7,19]. A visual summary of this filter's effect, kept vs. removed points, is provided in Figure 1.

**Figure 1: Effect of Density Filter on Training Dataset, Removed Low-Credibility 'Liquid' States**

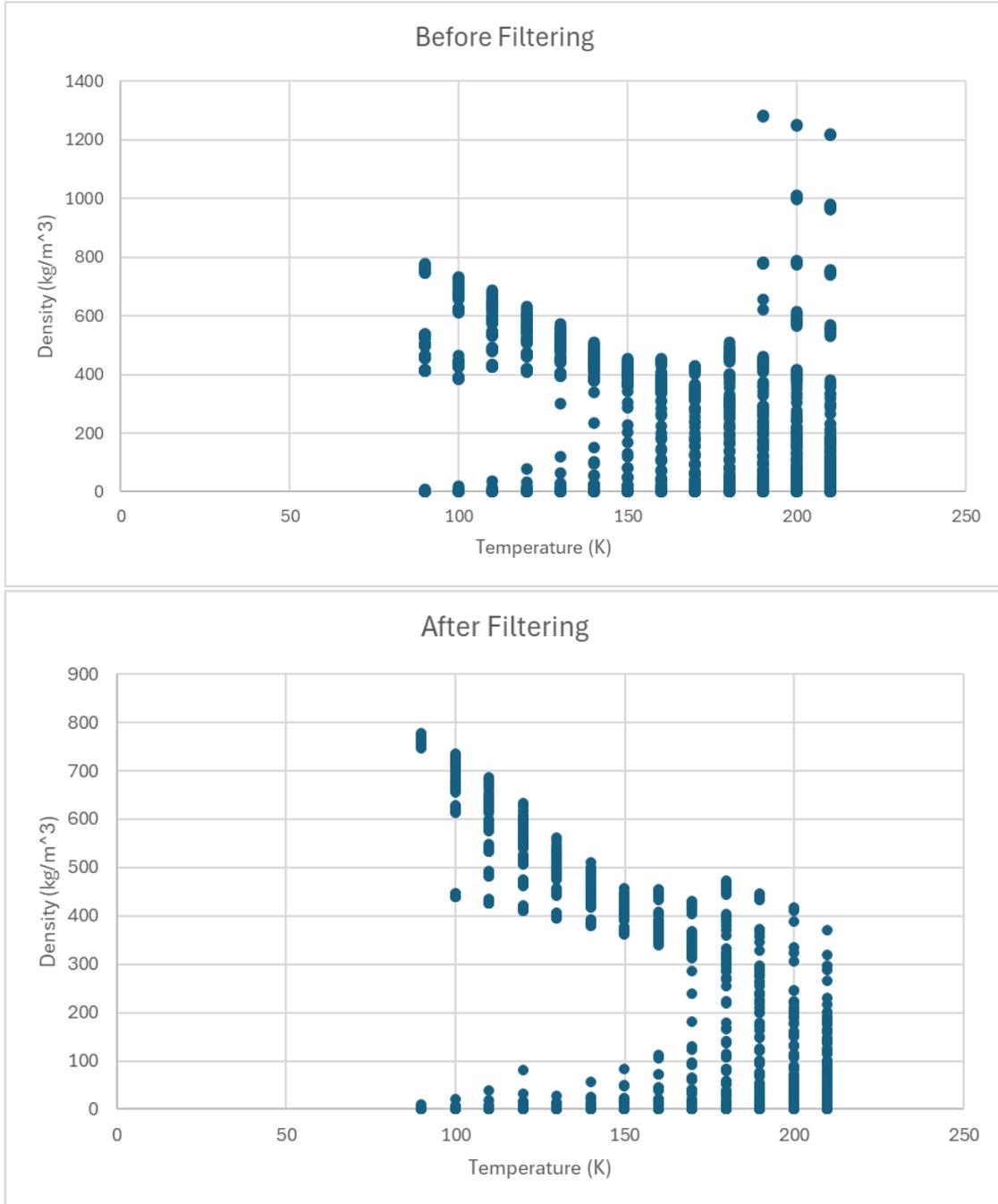

**Sampling coverage:** The curated set retains the intended breadth of the original design across the ternary triangle, enabling assessment of composition trends as well as binary and pure-component limits within the cryogenic band [7,20]. A ternary coverage plot is provided as Figure 2.

**Figure 2: Ternary Composition Coverage for the Curated Dataset**

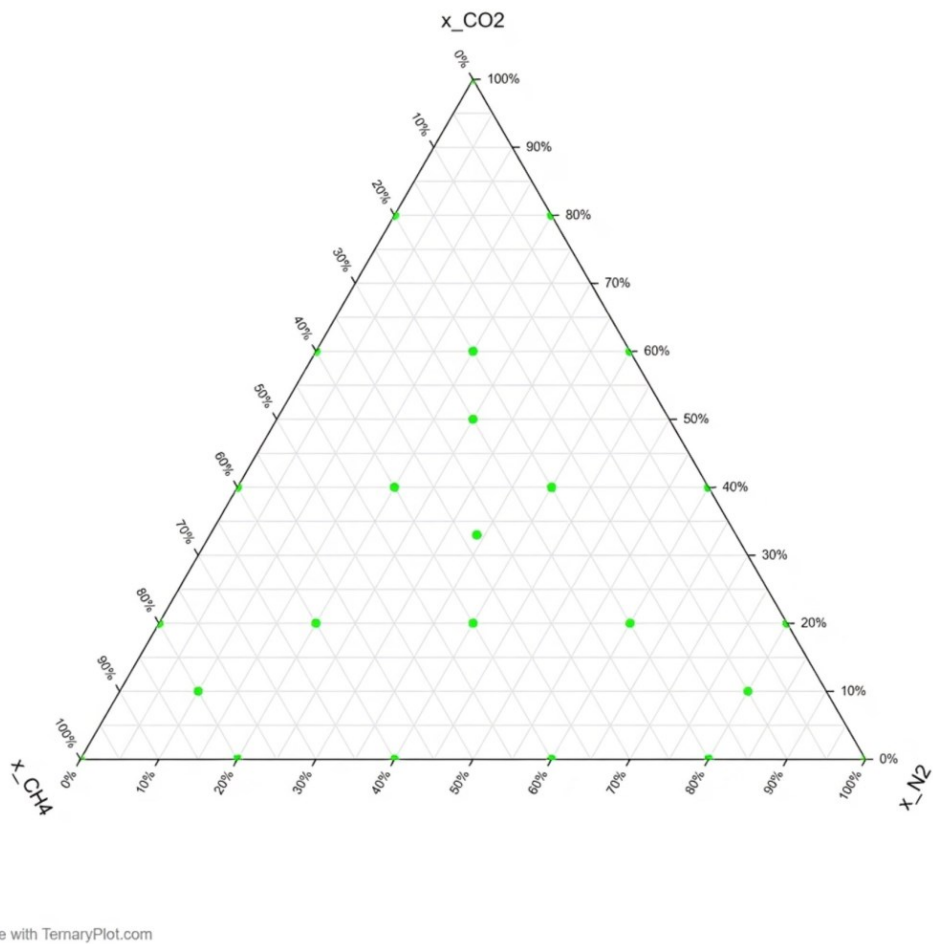

**Structural snapshots for graph inputs:** For each curated state, a short molecular-dynamics relaxation was performed using LAMMPS (29 Aug 2024 version) in a linux environment to obtain a representative low-energy structure. The resulting snapshots (e.g., `.xyz` files containing atomic numbers $z$ and 3D coordinates $pos$) serve as graph inputs to the model so that local geometry can inform predictions alongside thermodynamic state variables [21,22]. The structural generation is intentionally lightweight so that the dataset reflects physically plausible configurations without incurring the cost of long-horizon simulations at every point. For training/diagnostics, the study uses per-state MD snapshots; in the VLE solver, the study deploys a fixed template snapshot at inference.

**EoS-independent validation set:** In addition to the training/validation data derived from GERG-2008, this study prepared a separate set of 10 strategically chosen VLE points simulated with full molecular dynamics. This EoS-independent set is reserved for a final, external check on the surrogate's physical realism and for diagnosing whether observed errors align with regimes identified in single-phase tests [10,22].

**Summary:** In brief, the dataset comprises (i) an initial GERG-2008/CoolProp grid over 90-120 K, pressures up to 200 bar, and ternary compositions; (ii) a curated subset produced by a 15% density filter (from 5,200 to 1,516 states) intended to emphasize physically consistent single-phase behavior; (iii) per-state MD snapshots to provide structural features for graph learning; and (iv) a small, independent 10-point MD set for EoS-free assessment.

## 3. Model & Inputs

**Overview:** This study models thermophysical properties with a structure-aware graph neural network based on DimeNet++, using both thermodynamic state variables and per-state molecular snapshots as inputs and predicting residual properties as outputs [8,23]. A schematic of the data flow, from inputs to predicted residuals and final properties, is shown in Figure 3.

**Scalar (state) inputs:** Each example uses temperature $T$, base-10 log-transformed molar volume $logV_m$, and composition $x_i$ for $\{CO_2, CH_4, N_2\}$. (No $T_r$ or $\rho_r$ are used in the training forward pass.). All scalar inputs are taken directly from the curated GERG-2008/CoolProp dataset described in the Data section, with no additional targets introduced here [12,24].

**Structural (graph) inputs:** To expose local geometric cues that correlate with non-ideal mixture behavior, each curated state is paired with a short-relaxation molecular-dynamics snapshot. From these snapshots atomic numbers $z$ and 3D coordinates $pos$ are extracted, which define the molecular graph fed to DimeNet++. This pairing allows the model to combine composition-level information with instantaneous atomic neighborhoods, without incurring the cost of long MD trajectories during inference [8,25]. For training/diagnostics, the study uses per-state MD snapshots; in the VLE solver, the study deploys a fixed template snapshot at inference.

**Outputs and property reconstruction:** The network predicts residual Helmholtz energy $A_{res}$ only. Final properties $P$ and $U$ are obtained by differentiating $A_{res}$ (autograd) and adding ideal contributions, consistent with the residual/ideal decomposition used throughout the dataset. This residual-first formulation mirrors common practice in equation-of-state modeling and enables focused learning of non-ideal effects while preserving straightforward reconstruction of reportable quantities [26,27].

**Design rationale and scope:** The combination of $\{T, V_m, x_i, T_r, \rho_r\}$ with $\{z, pos\}$ is intended to balance physical interpretability and representational capacity: scalar features capture bulk state and composition, while graph features summarize short-range structure that influences departures from ideality at cryogenic conditions. This section focuses on the input-output contract of the surrogate; optimization details (e.g., losses and penalties) are described separately in the Training Protocol [28,29].

**Figure reference:** Figure 3 depicts: (i) scalar inputs $T, V_m, x_i, T_r, \rho_r$ (ii) structural inputs $z, pos$ derived from MD snapshots; (iii) the DimeNet++ blocks; and (iv) a single output head $A_{res}$ with downstream assembly into $P$ and $U$ (computed from $A_{res}$ plus ideal parts). This visual map anchors terminology used in the subsequent sections.

**Figure 3: Schematic of the GNN Surrogate**

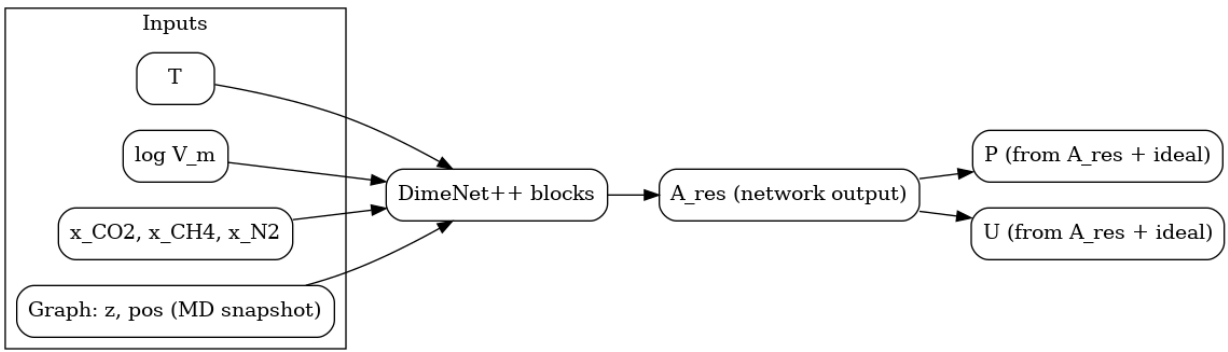

## 4. Training Protocol

**Overview:** This study employs a two-stage training schedule designed to first align the model with a physically meaningful energy landscape and then specialize it for the target residual properties [30,31]. Table 1 summarizes the configuration.

**Stage-1 (pretraining on residual Helmholtz energy):** The surrogate is first trained for 50 epochs to predict the residual Helmholtz energy $A_{res}$ using a Mean Squared Error (MSE) objective. Optimization uses Adam with a learning rate of 1e-4, batch size 1, and weight decay 1e-5. This stage provides a coarse alignment to the underlying thermodynamic surface before introducing derivative-sensitive targets [10,32].

**Stage-2 (fine-tuning on pressure):** Starting from Stage-1, the model is fine-tuned for 100 epochs using a Huber loss on pressure $P$ (computed from $A_{res}$ via autograd) plus a $C_v$ stability penalty (penalizes negative $C_v$). Optimization uses Adam (LR 1e-6, batch size 1, weight decay 1e-5) with a ReduceLROnPlateau scheduler. There is no separate $U$ data term and no explicit $\frac{dP}{dV}$ penalty in the implemented loss. No equations are introduced here; if EQ files are provided, the explicit penalty forms can be referenced [33,34,35].

**Reporting and reference table:** Table 1 lists for each stage: targets, epochs, loss components (MSE for $A_{res}$ in Stage-1; Huber($P$) + $C_v$ penalty in Stage-2), optimizer (Adam), learning rates (1e-4; 1e-6 with ReduceLROnPlateau), batch size (1), and weight decay (1e-5). The values above constitute the complete specification used in the experiments.

**Software environment:** Code was executed in Python 3.11 with the CUDA 12.1 runtime. Core deep-learning libraries were PyTorch 2.2.2+cu121 and TorchVision 0.17.2+cu121 (installed from the CUDA-12.1 wheel index). The PyTorch Geometric stack was installed from the torch-2.2.2+cu121 wheel index and comprised torch_geometric 2.6.1, torch_scatter 2.1.2+pt22cu121, torch_sparse 0.6.18+pt22cu121, torch_cluster 1.6.3+pt22cu121, and torch_spline_conv 1.2.2+pt22cu121. Numeric/data utilities were NumPy 1.26.4, pandas 2.2.2, and tqdm 4.67.1; thermophysical properties used CoolProp 7.0.0.

**Compute environment:** All experiments were run in Google Colab using the A-100 High-RAM GPU runtime. Session limits and memory constraints motivated choices such as batch size = 1 and the use of a template graph at inference; these practicalities provide context but do not alter the study's main findings.

**Table 1: Training Protocol and Hyperparameters for Both Stages**

| Stage | Target | Epochs | Loss Terms | Optimizer | Learning Rate | Batch Size | Weight Decay |
|---|---|---|---|---|---|---|---|
| 1 | $A_{res}$ (normalized) | 50 | MSE | Adam | 1e-4 | 1 | 1e-5 |
| 2 | Pressure (via $A_{res}$) | 100 | Huber($P$) + $C_V$ penalty | Adam | 1e-6 (+ ReduceLROnPlateau) | 1 | 1e-5 |

## 5. Surrogate Single-Phase Evaluation

**Scope:** This section evaluates pointwise single-phase property accuracy on held-out states; derivatives and VLE behavior are analyzed in Sections 7 and 8.

**Parity trends:** The parity plots for pressure and internal energy show tight clustering around the identity line for the majority of the validation points, indicating that the surrogate reproduces single-phase properties with high fidelity where the training distribution is well represented. Visually, the dominant pattern is a near-diagonal band with a small number of visible deviations that separate from the band. These outliers are sparse but distinct, and they foreshadow the heavy-tailed error behavior discussed below. Overall, Figure 4 communicates a clear dichotomy: broad agreement for most states and isolated departures that motivate targeted analysis. Note: In the diagnostic parity plots, pressure axes are labeled in bar for readability, while all computations and tables use SI (Pa).

**Figure 4: Parity Plots for P and U on the Validation Set**

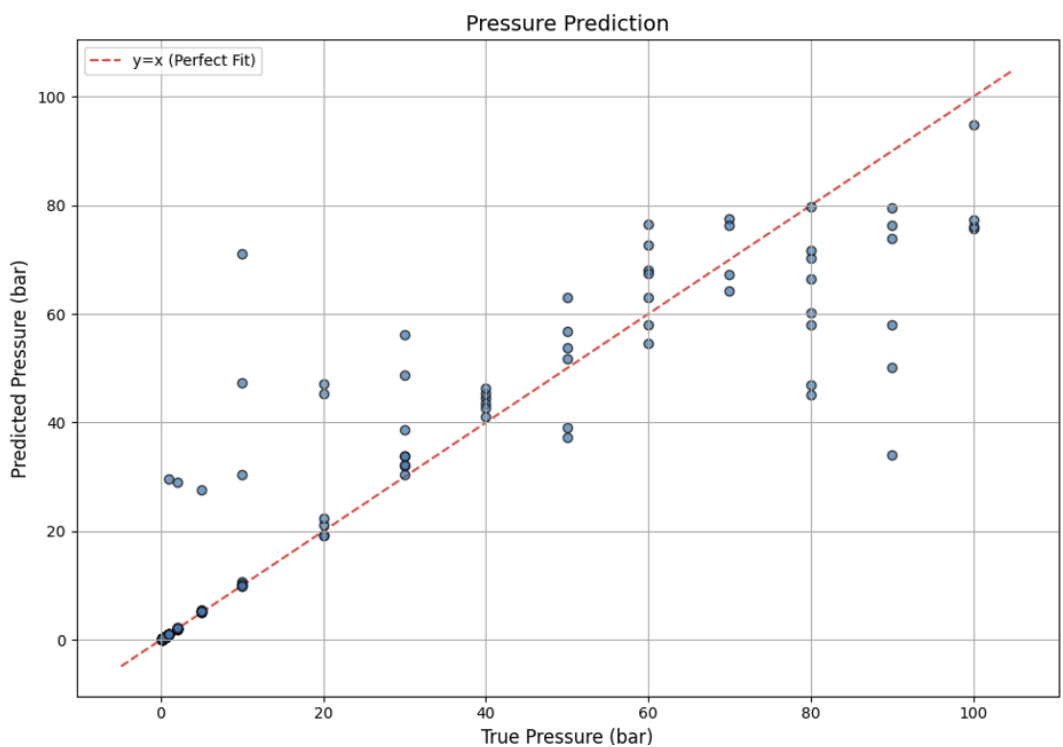

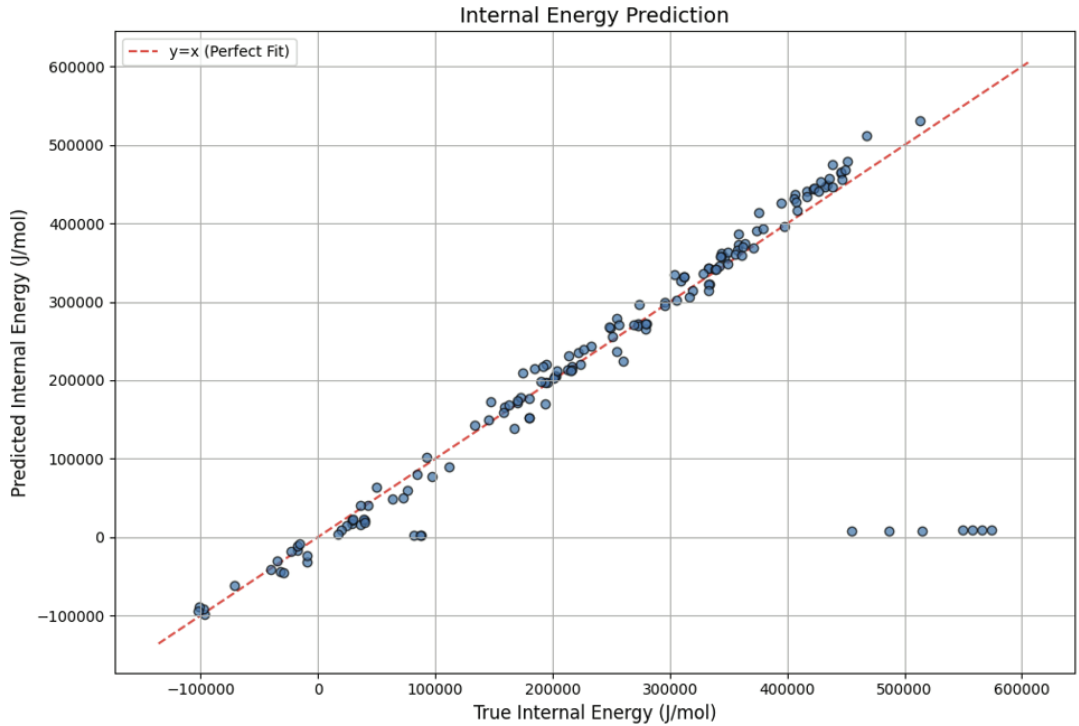

**Error distributions and median-mean divergence:** The histograms of percentage errors for pressure and internal energy reveal a pronounced difference between median and mean behavior. Quantitatively, the median absolute percentage errors are 5.68% for pressure and 5.68% for internal energy, while the mean absolute percentage errors are much larger (49.15% and 18.29%, respectively) consistent with a long right tail of difficult points. Root-mean-squared error values (12.31 bar for pressure; 113,739 J·mol$^{-1}$ for internal energy) further reflect the influence of these infrequent but high-magnitude deviations. Together, these diagnostics indicate that the model is robust for the bulk of cases but that a minority of states dominate the average error, a pattern that is visually consistent with the sparse outliers in Figure 4. Clarification: $U$ shown here is reconstructed from $A_{res}$; it is not a Stage-2 training target.

**Figure 5: Error Histograms for P and U, Showing Median vs. Mean**

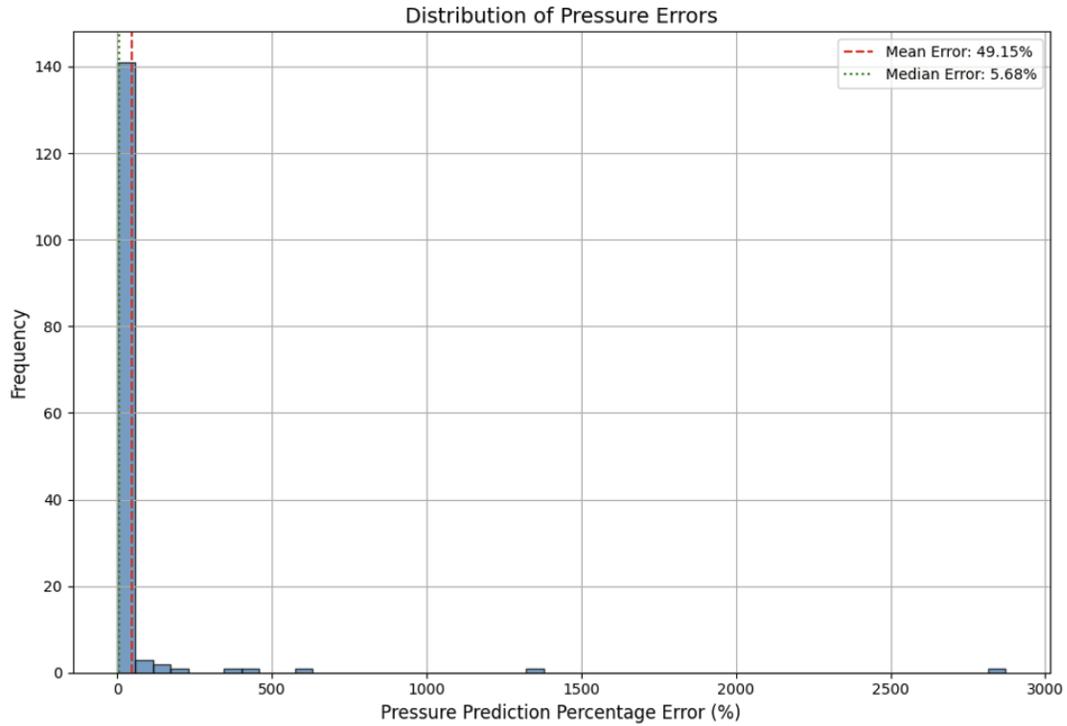

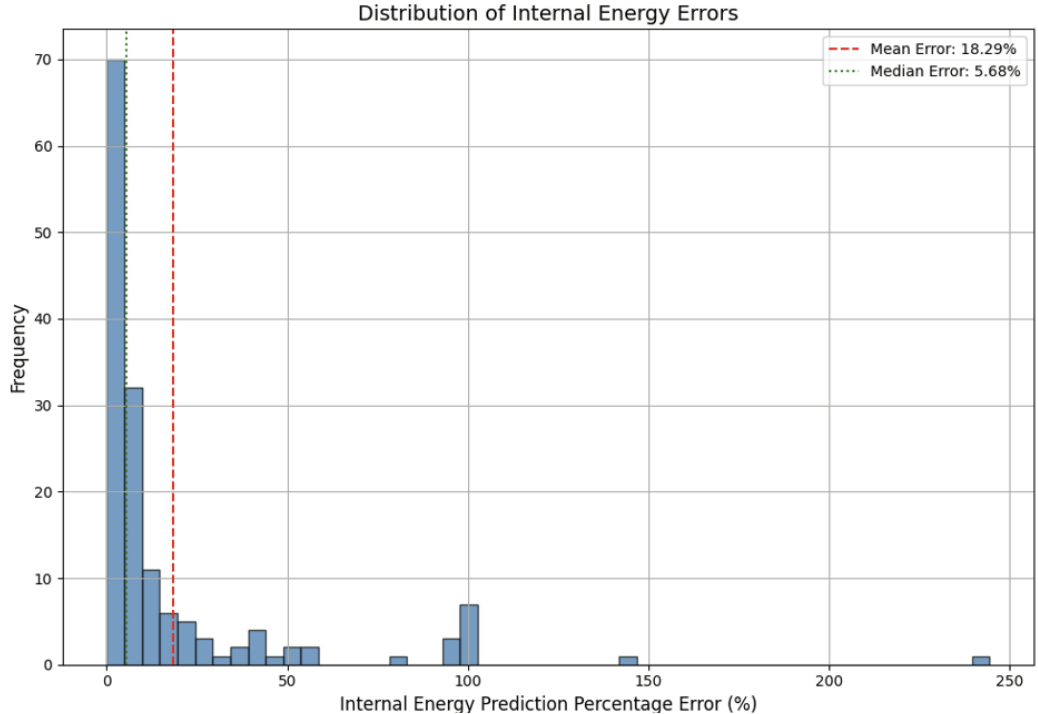

**Extremes and physical regimes:** The top-error cases listed in Table 2 concentrate in physically challenging regions aligned with dense, cold liquids. A representative example is nitrogen at 90 K and 774 kg·m⁻³, which appears among the highest-error points. This regime plausibly reflects strong short-range interactions and liquid structuring, consistent with where both empirical surrogates and classical models often struggle. Notably, these states are few in number relative to the full validation set, but their magnitude explains the gap between medians and means. The correspondence between Table 2 and the tails of Figure 5 reinforces a coherent picture: single-phase accuracy is strong over most of the cryogenic design space, and the largest discrepancies are concentrated in dense, cold-liquid conditions that lie near the edge of the model's effective interpolation envelope.

Table 2: Largest Absolute Percentage Errors and Corresponding State Points

| $T$ (K) | $\rho$ (kg/m³) | $x_{CO_2}$ | $x_{CH_4}$ | $x_{N_2}$ | $P_{true}$ (Pa) | $P_{pred}$ (Pa) | $P_{err}$ (%) |
|---|---|---|---|---|---|---|---|
| 90 | 774.485 | 0.0 | 0.0 | 1.0 | 500,000 | 2,764,522 | 452.90 |
| 90 | 751.427 | 0.0 | 0.0 | 1.0 | 8,000,000 | 4,514,332 | 43.57 |
| 100 | 0.562 | 0.0 | 0.4 | 0.6 | 8,000,000 | 3,861,012 | 28.70 |
| 110 | 483.163 | 0.0 | 0.8 | 0.2 | 3,000,000 | 3,389,392 | 12.98 |
| 90 | 747.580 | 0.0 | 0.0 | 1.0 | 4,000,000 | 4,454,172 | 11.35 |
| 100 | 438.935 | 0.0 | 1.0 | 0.0 | 7,000,000 | 7,748,472 | 10.69 |
| 110 | 428.238 | 0.0 | 1.0 | 0.0 | 20,000 | 18,029.52 | 9.85 |
| 110 | 0.458 | 0.0 | 0.6 | 0.4 | 10,000 | 10,646.27 | 6.46 |
| 100 | 668.399 | 0.0 | 0.2 | 0.8 | 50,000 | 52,910.30 | 5.82 |
| 110 | 573.940 | 0.0 | 0.4 | 0.6 | 20,000 | 19,023.12 | 4.88 |
| $T$ (K) | $\rho$ (kg/m³) | $x_{CO_2}$ | $x_{CH_4}$ | $x_{N_2}$ | $U_{true}$ (J/mol) | $U_{pred}$ (J/mol) | $U_{err}$ (%) |
| 90 | 747.580 | 0.0 | 0.0 | 1.0 | -9,342.143 | -32,185.016 | 244.51 |
| 90 | 751.427 | 0.0 | 0.0 | 1.0 | -32,007.845 | -43,569.922 | 36.12 |
| 90 | 774.485 | 0.0 | 0.0 | 1.0 | 28,598.995 | 21,296.625 | 25.53 |
| 100 | 0.562 | 0.0 | 0.4 | 0.6 | 97,417.075 | 77,493.461 | 20.45 |
| 90 | 771.524 | 0.0 | 0.0 | 1.0 | 191,583.880 | 217,847.203 | 13.71 |
| 100 | 438.935 | 0.0 | 1.0 | 0.0 | 35,737.857 | 40,600.676 | 13.61 |
| 110 | 0.893 | 0.0 | 1.0 | 0.0 | 406,847.112 | 436,379.000 | 7.26 |
| 100 | 668.399 | 0.0 | 0.2 | 0.8 | 248,896.012 | 266,172.438 | 6.94 |
| 110 | 0.255 | 0.0 | 0.4 | 0.6 | 421,688.170 | 443,241.219 | 5.11 |
| 110 | 483.163 | 0.0 | 0.8 | 0.2 | 189,738.880 | 198,597.391 | 4.67 |

**Takeaways for subsequent sections:** For downstream use, these results suggest that the surrogate can serve as a fast, accurate property estimator within the interior of the sampled single-phase region, while analyses that rely on precise behavior in dense, cold liquids warrant additional caution. This insight motivates the VLE feasibility study and the focused diagnostics that follow, where derivative quality rather than pointwise accuracy becomes the limiting factor.

## 6. VLE Solver Attempt: Fallback Audit

**Aim and setup:** The VLE driver is organized as a tiered routine. For each composition on a temperature-fixed grid, it first attempts a GNN-only solve: the surrogate provides residual thermodynamic quantities that the driver uses to seek a two-phase state satisfying equilibrium conditions (equality of phase pressures and component chemical potentials) [36]. If the GNN path cannot return stable, self-consistent values or the solver fails to converge, the routine falls back to CoolProp/GERG-2008 to retrieve the reference VLE for that state and records the status for auditability [6,17]. This fallback is thus not a shortcut for plotting; it is explicitly triggered after an unsuccessful GNN attempt [37].

**Cases evaluated:** Two binaries supported by the current solver wiring are tested; carbon dioxide/methane ($CO_2/CH_4$) and methane/nitrogen ($CH_4/N_2$); at 110 K and 120 K. For each (system, T) pair, the driver swept a uniform set of liquid-phase compositions across the bubble-line domain reported by CoolProp. The final Stage-2 checkpoint and the same template graph used in earlier experiments were employed; no changes were made to the training or solver hyperparameters between cases.

**Outcome - 0% GNN success, all accepted VLE points were CoolProp fallback, and the remaining attempted compositions were recorded as 'Solver Failed' (no VLE from CoolProp at that T/x or numerical failure):** Across all (system, T) cases and compositions probed, the driver accepted no GNN-produced equilibria. For compositions where a reference VLE value exists, the driver returned CoolProp Fallback; other attempted compositions were recorded as Solver Failed (no VLE returned at that $(T, x)$ or numerical failure on the reference path) [6,17].

**What "fallback" means here:** The driver always attempts the GNN path first; only after rejecting that attempt does it query CoolProp for a reference VLE. If CoolProp also cannot supply VLE at the candidate $(T, x)$, the row is logged as Solver Failed rather than plotted. This control flow ensures that "CoolProp Fallback" denotes a prior GNN failure, not a bypass [37].

**Figure 6: Evidence From Overlays (all accepted points are CoolProp fallback; missing points are "Solver Failed")**

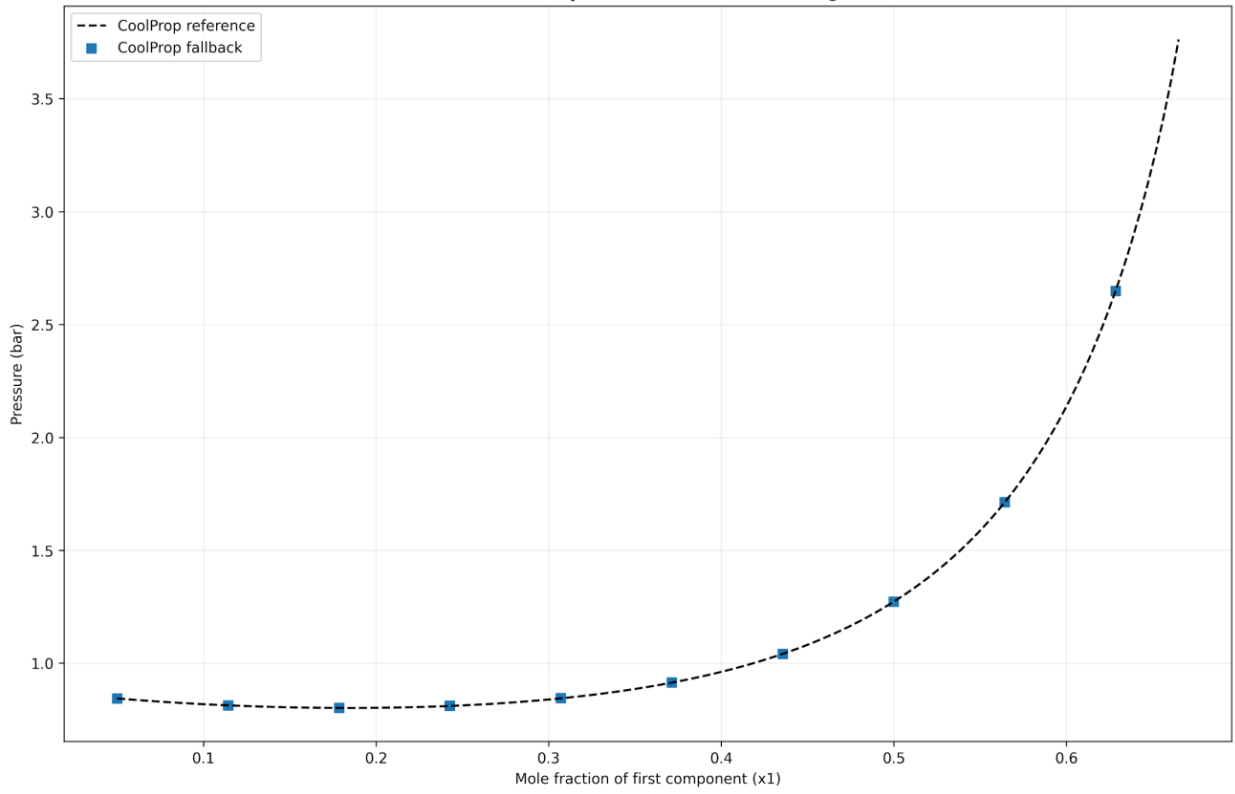
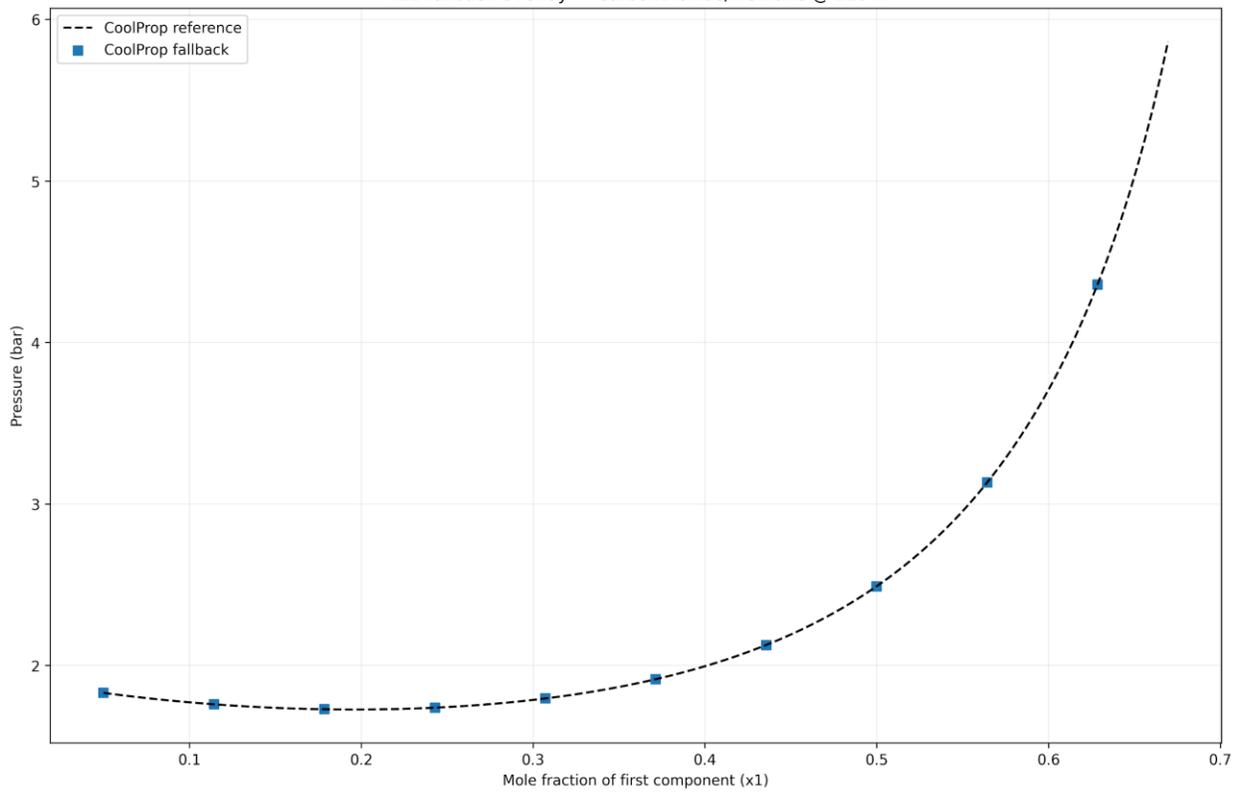

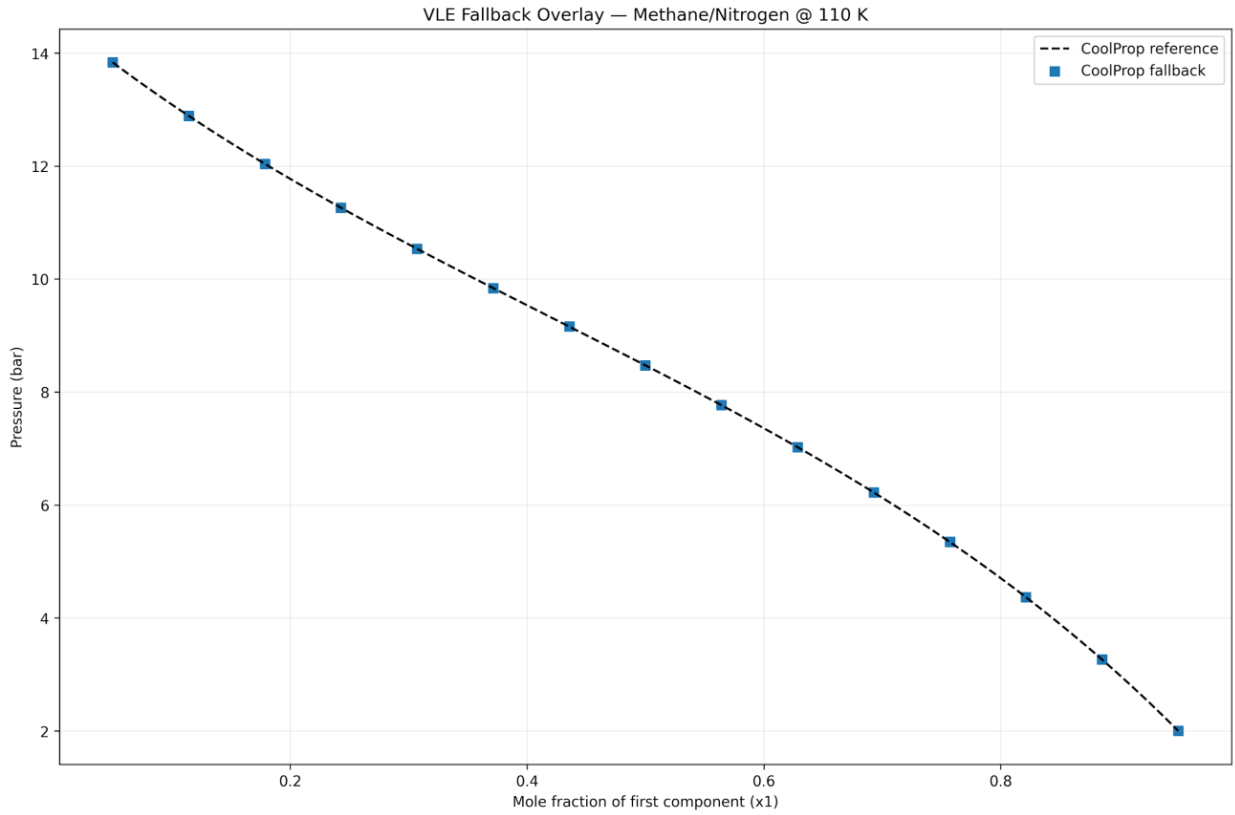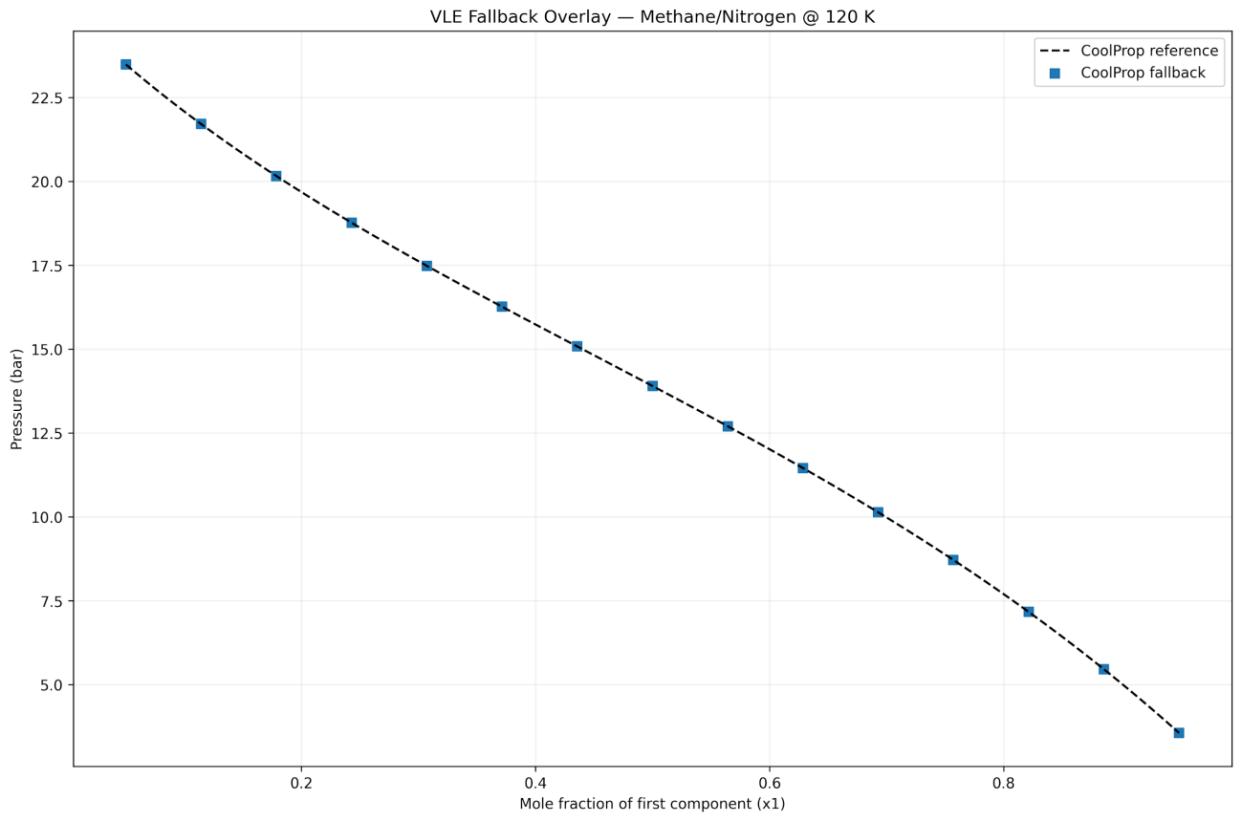

In Figure 6, markers denote accepted VLE points and therefore coincide with the CoolProp reference curves; no "GNN success" markers appear in any panel. Grid points recorded as Solver Failed have no VLE value to plot and thus do not appear in the overlays.

Table 3: VLE Fallback Audit (Per Case & Aggregated)

| T (K) | System | $P_{ref}$ (bar) | Status |
|---|---|---|---|
| 110 | Carbon Dioxide/Methane | 0.844 | CoolProp Fallback |
| 110 | Carbon Dioxide/Methane | 0.813 | CoolProp Fallback |
| 110 | Carbon Dioxide/Methane | 0.802 | CoolProp Fallback |
| 110 | Carbon Dioxide/Methane | 0.810 | CoolProp Fallback |
| 110 | Carbon Dioxide/Methane | 0.844 | CoolProp Fallback |
| 110 | Carbon Dioxide/Methane | 0.914 | CoolProp Fallback |
| 110 | Carbon Dioxide/Methane | 1.041 | CoolProp Fallback |
| 110 | Carbon Dioxide/Methane | 1.272 | CoolProp Fallback |
| 110 | Carbon Dioxide/Methane | 1.712 | CoolProp Fallback |
| 110 | Carbon Dioxide/Methane | 2.649 | CoolProp Fallback |
| 110 | Carbon Dioxide/Methane |  | Solver Failed |
| 110 | Carbon Dioxide/Methane |  | Solver Failed |
| 110 | Carbon Dioxide/Methane |  | Solver Failed |
| 110 | Carbon Dioxide/Methane |  | Solver Failed |
| 110 | Carbon Dioxide/Methane |  | Solver Failed |
| 120 | Carbon Dioxide/Methane | 1.830 | CoolProp Fallback |
| 120 | Carbon Dioxide/Methane | 1.760 | CoolProp Fallback |
| 120 | Carbon Dioxide/Methane | 1.729 | CoolProp Fallback |
| 120 | Carbon Dioxide/Methane | 1.739 | CoolProp Fallback |
| 120 | Carbon Dioxide/Methane | 1.796 | CoolProp Fallback |
| 120 | Carbon Dioxide/Methane | 1.915 | CoolProp Fallback |
| 120 | Carbon Dioxide/Methane | 2.127 | CoolProp Fallback |
| 120 | Carbon Dioxide/Methane | 2.490 | CoolProp Fallback |
| 120 | Carbon Dioxide/Methane | 3.132 | CoolProp Fallback |
| 120 | Carbon Dioxide/Methane | 4.359 | CoolProp Fallback |
| 120 | Carbon Dioxide/Methane |  | Solver Failed |
| 120 | Carbon Dioxide/Methane | | Solver Failed |
| 120 | Carbon Dioxide/Methane | | Solver Failed |
| 120 | Carbon Dioxide/Methane | | Solver Failed |
| 120 | Carbon Dioxide/Methane | | Solver Failed |
| 110 | Methane/Nitrogen | 13.833 | CoolProp Fallback |
| 110 | Methane/Nitrogen | 12.887 | CoolProp Fallback |
| 110 | Methane/Nitrogen | 12.038 | CoolProp Fallback |
| 110 | Methane/Nitrogen | 11.260 | CoolProp Fallback |
| 110 | Methane/Nitrogen | 10.534 | CoolProp Fallback |
| 110 | Methane/Nitrogen | 9.838 | CoolProp Fallback |
| 110 | Methane/Nitrogen | 9.156 | CoolProp Fallback |
| 110 | Methane/Nitrogen | 8.471 | CoolProp Fallback |
| 110 | Methane/Nitrogen | 7.765 | CoolProp Fallback |
| 110 | Methane/Nitrogen | 7.021 | CoolProp Fallback |
| 110 | Methane/Nitrogen | 6.220 | CoolProp Fallback |
| 110 | Methane/Nitrogen | 5.343 | CoolProp Fallback |
| 110 | Methane/Nitrogen | 4.366 | CoolProp Fallback |
| 110 | Methane/Nitrogen | 3.261 | CoolProp Fallback |
| 110 | Methane/Nitrogen | 2.000 | CoolProp Fallback |
| 120 | Methane/Nitrogen | 23.490 | CoolProp Fallback |
| 120 | Methane/Nitrogen | 21.712 | CoolProp Fallback |
| 120 | Methane/Nitrogen | 20.164 | CoolProp Fallback |
| 120 | Methane/Nitrogen | 18.772 | CoolProp Fallback |
| 120 | Methane/Nitrogen | 17.486 | CoolProp Fallback |
| 120 | Methane/Nitrogen | 16.267 | CoolProp Fallback |
| 120 | Methane/Nitrogen | 15.083 | CoolProp Fallback |
| 120 | Methane/Nitrogen | 13.905 | CoolProp Fallback |
| 120 | Methane/Nitrogen | 12.705 | CoolProp Fallback |
| 120 | Methane/Nitrogen | 11.458 | CoolProp Fallback |
| 120 | Methane/Nitrogen | 10.138 | CoolProp Fallback |
| 120 | Methane/Nitrogen | 8.718 | CoolProp Fallback |

| 120 | Methane/Nitrogen | 7.170 | CoolProp Fallback |
| 120 | Methane/Nitrogen | 5.462 | CoolProp Fallback |
| 120 | Methane/Nitrogen | 3.561 | CoolProp Fallback |

Table 3 provides the per-case audit log for every composition attempted by the driver, with outcomes labeled CoolProp Fallback or Solver Failed; there are no rows labeled "GNN Success."

**Interpretation:** The fallback audit provides a decisive answer to the feasibility question posed at the outset of the study: with the present architecture, data regime, and training protocol, the GNN surface did not support any VLE solves on the tested cryogenic binaries. Importantly, the audit also clarifies that the absence of GNN results is not due to bypassing the learned model; rather, the code attempted the GNN path first and only then invoked the reference implementation when the attempt failed [37]. As a consequence, the figures in this section should be read purely as reference VLE curves (for context and visualization), not as evidence of GNN VLE accuracy.

**Relation to diagnostic analysis:** The all-fallback outcome points toward deficiencies that manifest before or during equilibrium solves, e.g., insufficient smoothness or stability of GNN-derived derivatives near dense/cold regions where phase behavior is most sensitive. Section 7 investigates these issues without invoking a solver by examining pathwise smoothness and stability proxies on the learned surface [34,35,38].

## 7. Derivative-Quality Diagnostics

**Rationale:** Section 6 showed that the VLE driver never accepted a GNN-produced equilibrium: every attempted point fell back to the CoolProp/GERG-2008 reference. To understand why the solver fails without relying on the solver itself, the learned surface is probed with two lightweight diagnostics: (i) pathwise smoothness of the predicted pressure along local $logV_m$ paths at fixed $(T, x)$, and (ii) local stability flags derived from finite-difference slopes in small neighborhoods. These checks target the thermodynamic regularity conditions that practical VLE routines implicitly rely on, namely, monotonicity of pressure with respect to volume at constant temperature (mechanical stability) and non-negative isochoric heat capacity (thermal stability) [34,35,38].

**Figure 7: Pathwise Smoothness at Fixed** $(T, x)$ **(**$P$ **vs** $logV_m$ **With** $\frac{dP}{dlogV_m}$**)**

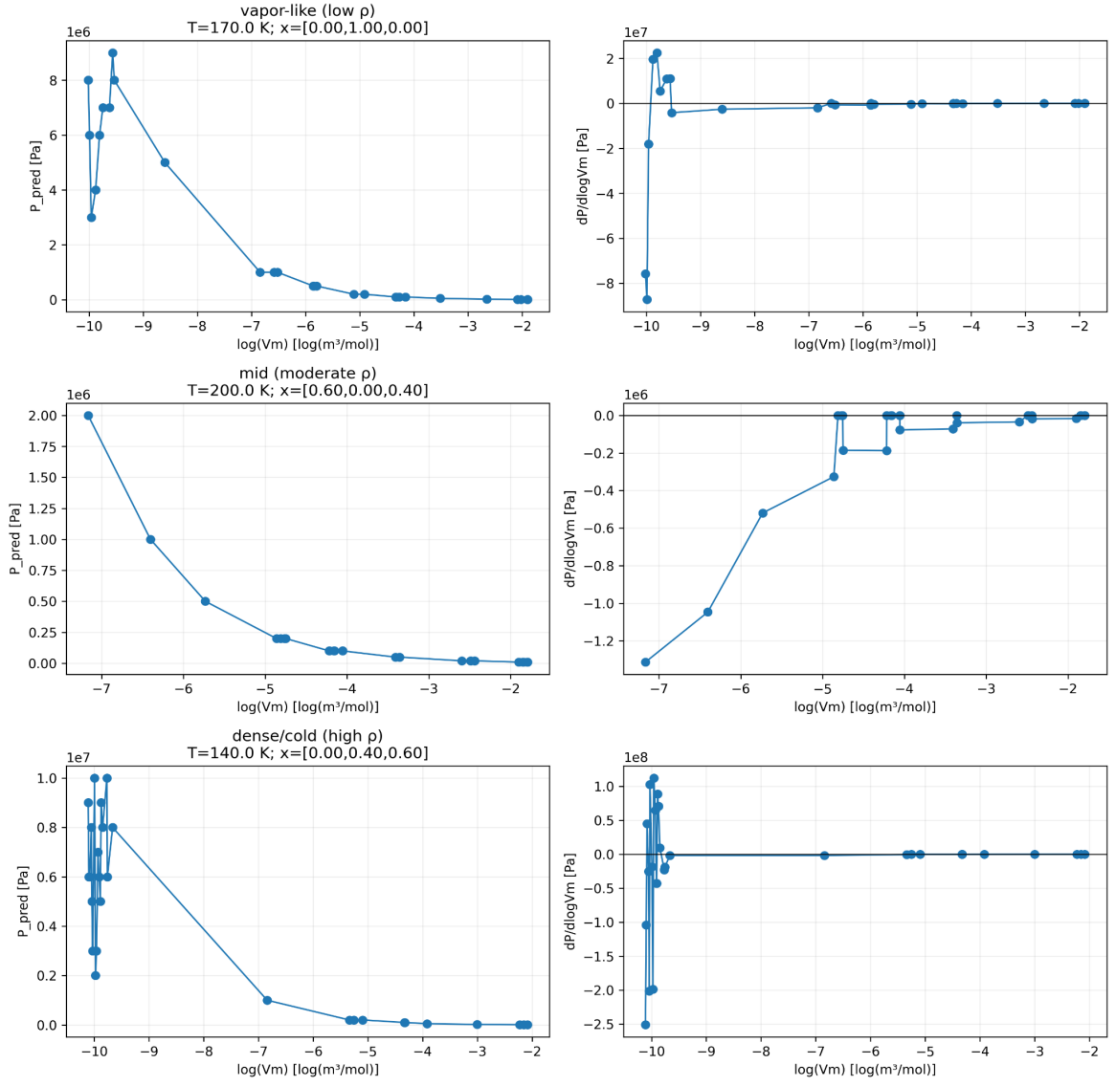

Three representative "anchors" are selected by density quantiles: vapor-like (low $\rho$), mid-regime, and dense/cold (high $\rho$). For each anchor a local neighborhood is gathered in $(T, x)$ using a K-nearest-neighbors rule, order those points by $logV_m$, and plot the surrogate's $P$ vs. $logV_m$ together with a centered finite-difference estimate of $\frac{dP}{dlogV_m}$. This procedure does not invoke equilibrium relations; it simply examines the shape and slope of the learned pressure surface along a physically meaningful coordinate.

As shown in Figure 7:
- Interior (vapor-like and mid-regime) neighborhoods produce smoother, largely monotone trends: $P$ decreases steadily as $logV_m$ increases, and the finite-difference slopes remain negative with small variation.
- Dense/cold neighborhoods display jagged trajectories and slope sign flips. In these panels the $P - logV_m$ curve exhibits kinks and local oscillations, and the estimated $\frac{dP}{dlogV_m}$ intermittently crosses zero.

This qualitative picture is consistent with a surface that interpolates well in most single-phase interior regions yet lacks the smoothness and directional consistency that robust VLE solvers expect near high-density, low-temperature states. The solver's line searches and root-finding steps depend on predictable slope behavior; irregular or sign-changing local slopes frustrate those procedures [34,35,38].

**Figure 8: Local Stability Rates By $logV_m$ (Mechanical & Thermal)**

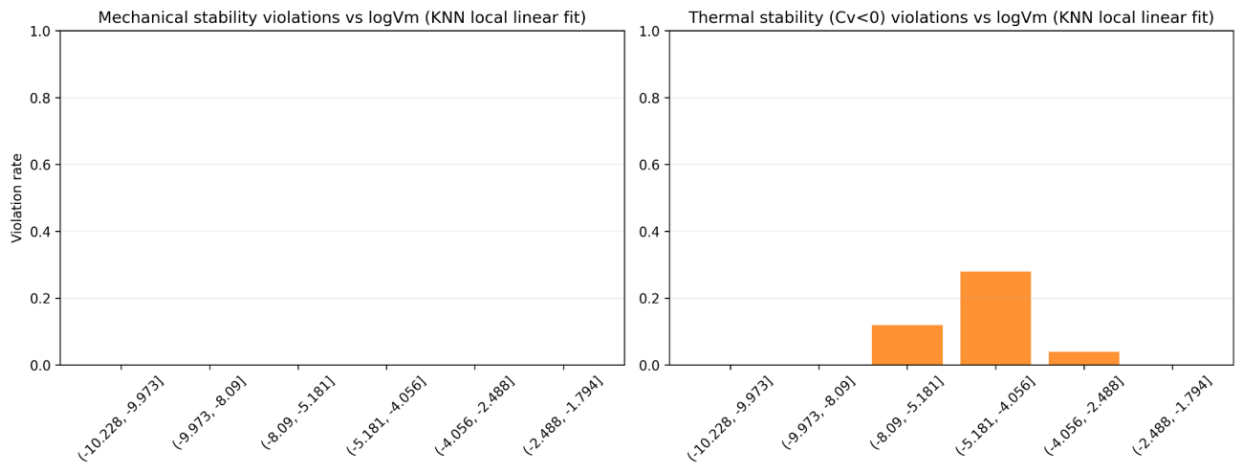

To quantify how often local regularity is violated, the validation states are binned by $logV_m$ (equal-frequency bins) and, within each bin, small KNN neighborhoods are fitted with simple finite-difference surrogates:
- Mechanical check: flag a violation when the local linear fit of $P$ vs. $logV_m$ yields a non-negative slope (proxy for $(\frac{\partial P}{\partial V})_T <= 0$).
- Thermal check: flag a violation when the local linear fit of $U$ vs. $T$ produces a negative temperature slope (proxy for $C_v >= 0$).

The counts and rates are then reported (no additional metrics).

Figure 8 summarizes the outcomes across $logV_m$ bins. In the run:
- Mechanical stability: the violation rate is 0.0 in every reported bin.
- Thermal stability: non-zero violation rates occur in several mid-to-high density bins (lower $logV_m$), specifically 12% in [-8.09, -5.181], 28% in [-5.181, -4.056], and 4% in [-4.056, -2.488]; the remaining bins report 0%.

The absence of mechanical-slope flags under this local test suggests the most salient issue is thermal irregularity (negative $C_v$ proxies) concentrated near the dense/cold regime. Together with the pathwise jaggedness in Figure 7, these findings indicate that the learned surface lacks the derivative quality required for reliable two-phase computation, even when pointwise single-phase errors are modest elsewhere [34,35,38].

These instabilities and slope inconsistencies provide a plausible explanation for the CoolProp fallbacks documented in Section 6: in regions where equilibrium calculations are most sensitive, the surrogate's local behavior is insufficiently smooth or thermodynamically consistent for the solver to converge.

## 8. Single-Phase Inference Latency

**Scope:** Per-call latency is benchmarked on the same 152 single-phase validation states analyzed in Section 5. Accuracy for these states is discussed in Section 6; here the study reports compute cost only. CoolProp/GERG-2008 serves as the reference baseline [6,17]. Early trials used an alternative flash path that emitted 'does not support mixtures' messages; the benchmark here uses HEOS (Helmholtz-Energy Equation of State) with DmassT_INPUTS and umolar(), and the n values in Table 4 confirm successful calls for all benchmarked states.

**Setup:** For each state $(T, \rho, x)$, the study times (i) the GNN inference path used in Section 5 and (ii) CoolProp on the same rows ("GNN-OK"). As an additional baseline, the study also times CoolProp across all 152 validation states ("All-152"). The reported numbers come directly from Table 4; Figure 9 shows the corresponding latency cumulative distribution functions (CDFs).

Table 4: Latency Summary (GNN vs CoolProp)

| Engine | Subset | n | Median (ms) | p95 (ms) | Mean (ms) |
|---|---|---|---|---|---|
| GNN | GNN-OK | 152 | 35.865 | 73.695 | 43.738 |
| CoolProp | GNN-OK | 152 | 0.057 | 0.070 | 0.059 |
| CoolProp | All-152 | 152 | 0.056 | 0.065 | 0.057 |

**Qualitative view (Figure 9):** The CoolProp curves are essentially collapsed near the origin (sub-millisecond scale), while the GNN CDF rises over tens of milliseconds, with a noticeable long tail. This visual separation mirrors the tabled medians and p95 values.

Figure 9: Latency CDF (GNN vs CoolProp)

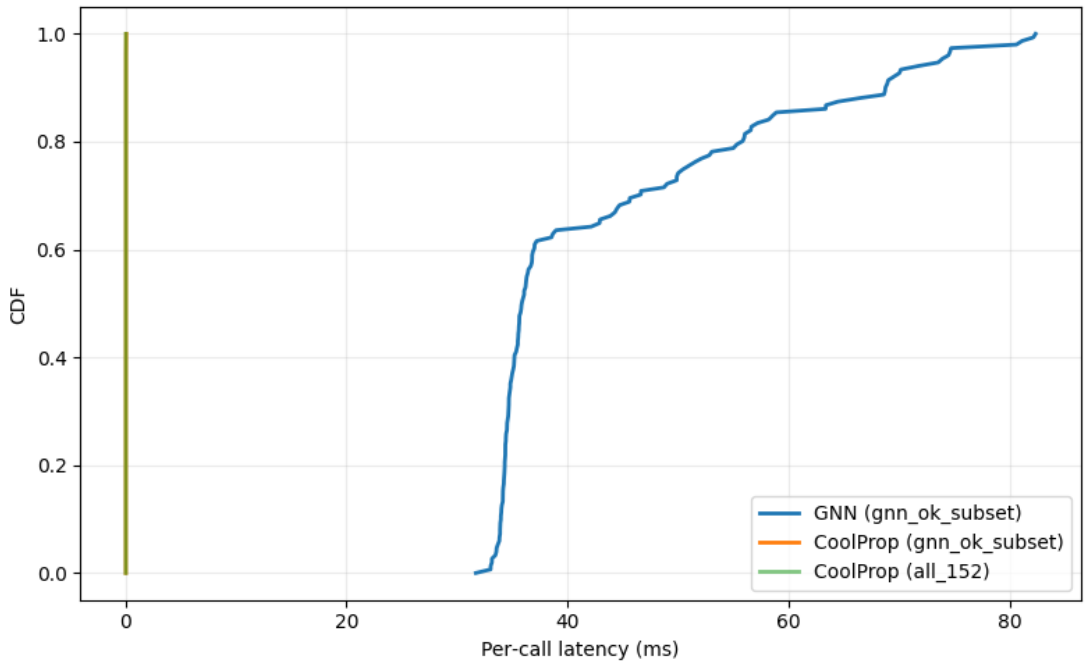

In this setup, the surrogate does not provide a single-phase speed advantage relative to the CoolProp/GERG-2008 baseline. The GNN timings reflect the end-to-end inference routine used in the diagnostics (including feature preparation and any per-state I/O), whereas CoolProp timings are direct property evaluations; the comparison reflects practical runtime of the actual pipelines employed. The latency evidence here is therefore best read as a practical runtime comparison of the actual pipelines used in Section 5 rather than a lower-bound on possible model inference speed.

## 9. Discussion

**Problem restated:** The aim was to learn a Helmholtz-energy surrogate for cryogenic $CO_2 - CH_4 - N_2$ mixtures and to assess whether that surrogate is sufficiently accurate and well-behaved to support VLE calculations. In parallel, its single-phase utility is also evaluated, both in terms of accuracy (interpolation on a held-out validation set) and practical runtime, relative to a GERG-2008 reference implementation.

**Limitations (resource context):** Resource limits (ephemeral Colab sessions, RAM limitations) influenced some design choices (e.g., batch size = 1, template snapshot at inference) but do not change the study's main findings that the surrogate is not VLE-ready and offers no single-phase speed advantage.

**What worked (single-phase interpolation):** The results in Section 5 indicate that, within its training regime, the model interpolates single-phase properties reasonably well. Parity plots and error distributions are consistent with a surrogate that captures broad trends, with systematic outliers appearing in dense/cold liquid conditions. These outliers are not random; they recur in similar regions of state space and composition, suggesting that data scarcity, representation limits, or loss-surface trade-offs produce a patterned degradation rather than incidental noise.

**What did not work (VLE):** Section 6 shows that this single-phase interpolation strength did not translate into a working equilibrium solver. On both binaries tested ($CO_2/CH_4$ and $CH_4/N_2$) and at two cryogenic temperatures per system, the driver achieved 0% GNN success and mostly fallback to the CoolProp/GERG-2008 reference. This outcome matters because VLE requires not only faithful pointwise values but also consistent and smooth derivatives that enable reliable root-finding and line searches across two phases [12,36,38]. In other words, single-phase accuracy is necessary but not sufficient for VLE.

**Why it did not work (derivative quality):** Section 7 provides solver-free evidence that helps explain the all-fallback outcome. Along local $logV_m$ paths at fixed $(T, x)$, the surrogate is smooth and monotone in interior regions but becomes jagged with slope sign flips in dense/cold neighborhoods. Aggregating local stability checks over the validation set reinforces this picture: while gross mechanical inversions are not prominent under the test, thermal-stability flags concentrate in mid-to-high-density bins near the liquid-like regime. Together, these findings identify a derivative-quality deficit, particularly around dense/cold states where VLE sensitivity is highest, which plausibly prevents a GNN-only equilibrium calculation from converging.

**Practicality (latency):** Section 8 further contextualizes the surrogate's single-phase practicality: when evaluated on the same validation states as Section 5, the current end-to-end inference path shows no speed advantage relative to the CoolProp/GERG-2008 baseline. This comparison reflects the pipelines as actually used in the study (including feature preparation and I/O on the GNN side), and it underscores that, absent VLE capability, the surrogate does not presently offer a runtime benefit for property queries either.

**What the negative results give us:** Although the original goal (a working VLE solver driven by a learned Helmholtz energy) was not achieved, the study draws clear boundaries for this approach and contributes concrete diagnostics for future work. The study pinpoints where and how the surface loses the regularity VLE needs (dense/cold regions; thermal-stability proxies; pathwise slope behavior), and the study provides a transparent fallback audit that others can reproduce. These deliverables clarify which aspects of data coverage, inductive bias, and training signals would need to change before a GNN surrogate can plausibly shoulder equilibrium calculations, and they establish an honest baseline against which such advances can be measured.

## 10. Conclusion & Future Work

**Conclusions:** This study assembled a curated cryogenic dataset for $CO_2 - CH_4 - N_2$ mixtures, trained a DimeNet++-based Helmholtz-energy surrogate with a two-stage protocol, and evaluated its behavior in single- and two-phase contexts. Within the covered regime, the surrogate delivers accurate single-phase interpolation (Section 5). However, the fallback audit shows that the VLE driver achieved zero GNN solutions and relied almost entirely on the CoolProp/GERG-2008 reference (Section 6). Solver-free diagnostics reveal derivative and stability issues concentrated in dense/cold regions (Section 7), and an end-to-end latency benchmark indicates no speed advantage over CoolProp for single-phase calls (Section 8). The study therefore concludes that, in its current form, the surrogate is not solver-ready for VLE and does not provide a practical runtime benefit for property queries. Its primary value is methodological: a transparent negative result with concrete diagnostics that delineate where this approach fails and how future work can target those gaps.

**Future work:**
- Incorporating physics-informed loss terms that directly enforce thermodynamic consistency during training.
- Training on larger, more diverse datasets that include points near the phase boundary.
- Exploring alternative solver algorithms that are less sensitive to small errors in derivatives.

**Availability: No public package accompanies this work; result logs and figure scripts are available on request.**